\documentclass[lettersize,journal]{IEEEtran}
\usepackage{amsmath,amsfonts}
\usepackage{array}
\usepackage[caption=false,font=normalsize,labelfont=sf,textfont=sf]{subfig}
\usepackage{textcomp}
\usepackage{stfloats}
\usepackage{url}
\usepackage{verbatim}
\usepackage{graphicx}
\usepackage{cite}

\usepackage[dvipsnames]{xcolor}
\usepackage[ruled,vlined,linesnumbered]{algorithm2e}
\usepackage{bbding}
\usepackage{svg}

\usepackage{algpseudocode}
\usepackage{pgfplots}
\usepackage{listings}
\usepackage{titlesec}
\usepackage{tikz}
\usetikzlibrary{arrows, positioning}
\usepackage{enumitem}   
\usepackage{multirow}
\usepackage{booktabs}
\usepackage{pifont} 
\usepackage{hyperref}
\usepackage{float}

\newcommand{\eat}[1]{}

\newcommand{\cmark}{\ding{51}}
\newcommand{\xmark}{\ding{55}}

\begin{document}

\title{Context-Aware Scientific Knowledge Extraction on Linked Open Data using Large Language Models}


\author{Sajratul Yakin Rubaiat,~\IEEEmembership{Member,~IEEE,}
        Hasan M Jamil,~\IEEEmembership{Member,~IEEE,}
\IEEEcompsocitemizethanks{\IEEEcompsocthanksitem S. Y. Rubaiat is with the Department of Computer Science, University of Idaho, Moscow, ID, USA (e-mail: ruba3062@vandals.uidaho.edu).\protect\\
\IEEEcompsocthanksitem H. M. Jamil is with the Department of Computer Science, University of Idaho, Moscow, ID, USA (e-mail: jamil@uidaho.edu).}
\thanks{Manuscript received January 1, 2025; revised January 1, 2025.}}


\IEEEpubid{0000--0000/00\$00.00~\copyright~2021 IEEE}

\maketitle

\begin{abstract}
The exponential growth of scientific literature presents a significant challenge for researchers seeking to extract and synthesize relevant knowledge. Traditional search engines often return a large number of sources without directly providing detailed answers, while general-purpose Large Language Models (LLMs) may offer concise responses that lack depth or fail to incorporate the most up-to-date information. Furthermore, LLMs with search capabilities are often limited by their context window, resulting in short, incomplete answers. This paper introduces WISE (\textbf{W}orkflow for \textbf{I}ntelligent \textbf{S}cientific Knowledge \textbf{E}xtraction), a novel system that addresses these limitations by combining LLMs with a structured, multi-layered workflow to extract, refine, and rank scientific knowledge tailored to specific queries. WISE employs an LLM-powered, tree-based architecture with a customized search function to iteratively refine extracted data, focusing on query-aligned and context-aware information while actively avoiding redundancy. Dynamic scoring and ranking mechanisms prioritize unique contributions from each source, and adaptive stopping criteria minimize processing overhead. WISE delivers detailed, well-organized, and highly informative answers by systematically exploring and synthesizing knowledge from diverse sources. Experiments focused on biological queries related to \textit{HBB} gene-associated diseases demonstrate that WISE reduces the volume of processed text by over 80\% while simultaneously achieving significantly higher recall compared to baseline methods, including leading search engines and other LLM-based approaches. Further analysis using ROUGE and BLEU metrics reveals that WISE's output is more unique compared to other systems, and a novel level-based evaluation metric shows that WISE provides more in-depth information. This paper also explores how the WISE workflow can be adapted as a general framework for diverse research domains, such as \textit{drug discovery, material science, and social science}, enabling efficient knowledge extraction and synthesis from unstructured scientific papers and web sources across a wide array of research domains.
\end{abstract}

\begin{IEEEkeywords}
Knowledge Discovery, Large Language Models, Information Filtering, Scientific Data Extraction, Knowledge Enrichment, Gene-Disease Associations
\end{IEEEkeywords}

\section{Introduction}

\IEEEPARstart{T}{he} relentless expansion of scientific knowledge, reflected in the ever-increasing volume of published literature, presents a formidable challenge for researchers seeking to extract, synthesize, and contextualize relevant information \cite{10.1371/journal.pdig.0000347, doi:10.4137/BII.S31559}. While traditional search engines and general-purpose Large Language Models (LLMs) offer some assistance, they often fall short in providing domain-specific insights, filtering irrelevant content, and efficiently managing the sheer volume of data \cite{zhai2024large, 10.1145/3626772.3657733, ziems2023large}. Traditional search engines typically return a large number of sources, requiring users to manually sift through them to extract relevant information, rather than providing direct, synthesized answers. Navigating this complex information ecosystem manually is both labor-intensive and error-prone, with researchers facing the risk of overlooking critical details as the volume of data grows exponentially.

Consider, for example, a researcher investigating gene-disease associations related to the \emph{HBB} gene. Starting from a single authoritative source like the HGNC \cite{hgnc4827}, they might encounter 24 relevant sources. Exploring just one of these, such as ClinVar \cite{clinicalgenome_hgnc4827}, could unveil hundreds more sources, leading to a rapidly expanding tree of interconnected resources, exemplified by platforms like NCBI \cite{ncbi_nbk1435}. This exponential growth of linked resources quickly overwhelms traditional search systems, making it difficult to extract pertinent insights efficiently. Even experienced investigators struggle to filter out superfluous data and focus on the most valuable information, a challenge amplified for newcomers. Consequently, critical information may be missed, and the time required to gain a complete, integrated understanding escalates dramatically.

\IEEEpubidadjcol

Purely automated approaches also encounter significant difficulties in this context \cite{garcia2024reviewscientificknowledgeextraction, systems11070351, 10020725}. The sheer volume of interconnected data can lead to computationally expensive and strategically ineffective processes without robust mechanisms for pruning irrelevant or redundant content. A key challenge lies in determining when to stop searching; continued exploration without clear stopping criteria often yields diminishing returns, underscoring the need for a balanced workflow that ensures thorough yet efficient exploration while prioritizing high-value information \cite{sneyd-stevenson-2019-modelling, 10.1007/978-3-031-75434-0_23}.

\begin{table*}[h]
    \centering
    \caption{System Comparison}
    \begin{tabular}{p{4cm}ccccc}
        \hline
        \textbf{Feature} & \textbf{WISE} & \textbf{ChatGPT} & \textbf{ChatGPT with Search} & \textbf{Gemini} & \textbf{Google Search} \\
        \hline
        Number of Diseases Identified & 16 & 9 & 7 & 2 & 3 \\
        Recall  & 0.84 & 0.47 & 0.36 & 0.10 & 0.15 \\
        Average Level of Detail \textsuperscript{a} & 3.8 & 3.33 & 3.42 & 2.5 & 3.0 \\
        Structured Output & \cmark & \cmark & \cmark & \cmark & \xmark \\
        Inclusion of Sub-variations & \cmark & \xmark & \cmark & \xmark & \xmark \\
        Source Citation & \cmark & \xmark & \cmark & \xmark & \cmark \\
        Identification of Rare Conditions & \cmark & \xmark & \xmark & \xmark & \xmark \\
        Up-to-Date Information & \cmark & \xmark & \cmark & \xmark & \cmark \\
        \hline
    \end{tabular}
    \label{tab:comparison-table}
    \footnotesize

    \textsuperscript{a} Average Level of Detail: Represents the average depth of information provided across all diseases identified by the system, based on a 0-5 scale where higher values indicate more detailed information (see Section \ref{experiment} for level criteria). \\
\end{table*}

\begin{figure}[h]
\centering
 \includegraphics[width=1\linewidth]{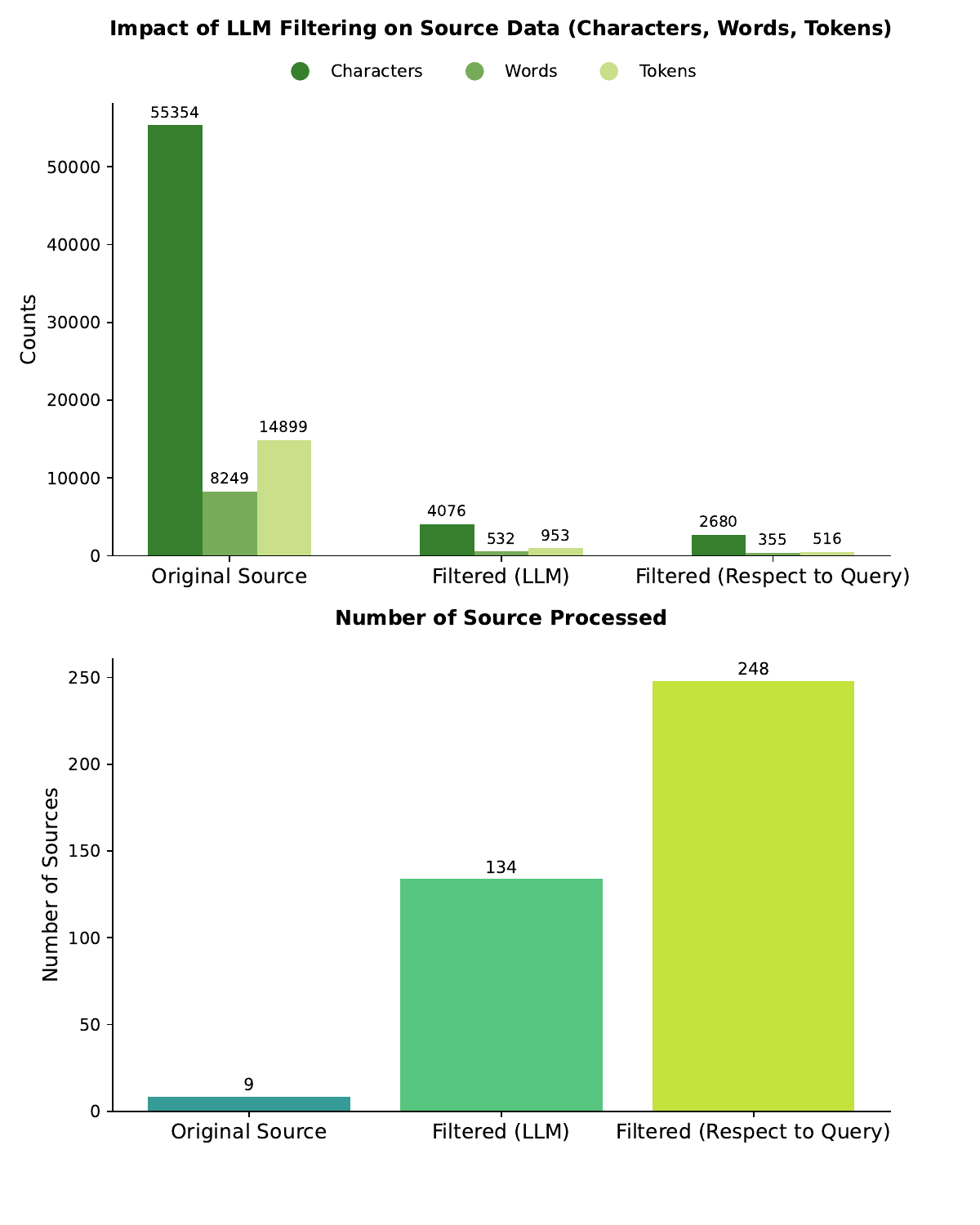}
\caption{Number of sources, such as UniProt \cite{uniprotP68871}, that can be processed simultaneously for ranking by advanced LLMs like GPT-4o demonstrate significant improvements when applying LLM-based filtering with and without query-specific relevance. Given that GPT-4o supports a 128,000-token context window, the number of websites that can be processed is calculated as: Number of websites = 128,000 / Tokens per Website}
\label{impact_of_LLM_filtering}
\end{figure}

While Large Language Models (LLMs) show promise in specific aspects of information retrieval \cite{vaswani2023attentionneed}, their inherent limitations hinder their ability to fully address the scale of these challenges. General-purpose LLMs, often provide concise answers that lack the depth and detail required for complex scientific queries, and may not incorporate the most recent findings. For instance, state-of-the-art models like GPT-4o \cite{openai2024gpt4o}, despite having a context window of \textit{128000} tokens, are practically limited to processing data from only about eight sources simultaneously, such as UniProt \cite{uniprotP68871}, as illustrated in Figure~\ref{impact_of_LLM_filtering}. Although capable of ranking and comparing content within this limited scope, LLMs are constrained by their narrow context window, further reduced by factors like search history. In specialized domains such as biology and medicine, where nuanced and detailed insights are crucial, relying solely on LLM-based searches proves insufficient. These challenges highlight the critical need for combining the capabilities of LLMs with strategic workflows to achieve comprehensive and efficient knowledge retrieval.

To address these challenges, we introduce \textbf{WISE} (\textbf{W}orkflow for \textbf{I}ntelligent \textbf{S}cientific \textbf{E}xtraction), a novel, scalable, tree-based framework that integrates LLM-driven filtering, dynamic ranking, and adaptive stopping criteria. WISE is designed to deliver detailed, well-organized, and highly informative answers by systematically exploring and synthesizing knowledge from diverse sources. WISE mirrors the approach of a diligent researcher: identifying relevant information, discarding duplicates, exploring promising leads, and recognizing when further pursuit yields diminishing returns. WISE begins by employing LLMs to filter large text corpora based on a user's domain-specific query, ensuring that only contextually relevant and manageable segments proceed to subsequent stages. It then assigns scores to extracted sources, quantifying their unique contributions relative to previously processed material, thereby minimizing redundancy and focusing on content that adds genuine value. This dynamic ranking process prioritizes high-value information while pruning low-impact paths. The iterative refinement continues layer by layer, with WISE's knowledge container—the growing repository of extracted, query-specific insights—expanding until incremental findings diminish (Figure \ref{data_extraction_status_across_categories}). At this point, WISE intelligently halts further searches, conserving computational resources while delivering comprehensive, contextually nuanced results. In essence, WISE achieves a balance between breadth and depth, effectively leveraging the strengths of LLMs while mitigating their limitations through intelligent pruning and carefully considered stopping criteria.

Our key \textbf{contributions} are summarized as follows:

\begin{enumerate}
    \item \textbf{Scalable, Tree-Based Architecture:} We introduce a novel, tree-structured workflow (Section~\ref{system_design}) that efficiently navigates large, heterogeneous datasets. This architecture leverages LLM-based filtering at each layer to incrementally refine data subsets according to domain-specific queries, ensuring scalability and focus.

    \item \textbf{Dynamic Ranking and Pruning:} We present a transparent scoring mechanism (Sections~\ref{system_design} and \ref{experiment}) that quantifies each source's unique knowledge contribution. By dynamically ranking sources based on their added value and employing intelligent pruning, WISE focuses computational resources on the most promising leads, effectively filtering out redundant or low-value content.

    \item \textbf{Adaptive, Expert-Inspired Exploration:} Our approach (Sections~\ref{system_design} and \ref{experiment}) mirrors expert-driven inquiry by progressively deepening the search along promising paths while adaptively halting exploration when further gains are minimal. This ensures a balanced blend of breadth and depth, optimizing both the efficiency and effectiveness of the knowledge discovery process.

    \item \textbf{Demonstrated Effectiveness in Gene-Disease Association Discovery:} Through empirical evaluation on gene-disease association queries (Section~\ref{results}), we demonstrate that WISE significantly outperforms baseline methods, including traditional search engines and general-purpose LLMs, in terms of recall, uniqueness of extracted information (ROUGE/BLEU), and depth of knowledge (level-based analysis).

    \item \textbf{Versatile Applications:} We showcase WISE's adaptability and potential impact through diverse applications (Section~\ref{applications}), including drug discovery, material science, and social science, highlighting its ability to generalize across a wide range of research domains.
\end{enumerate}

\section{System Design}
\label{system_design}

\begin{figure}[h]
 \centering
 \includegraphics[width=1\linewidth]{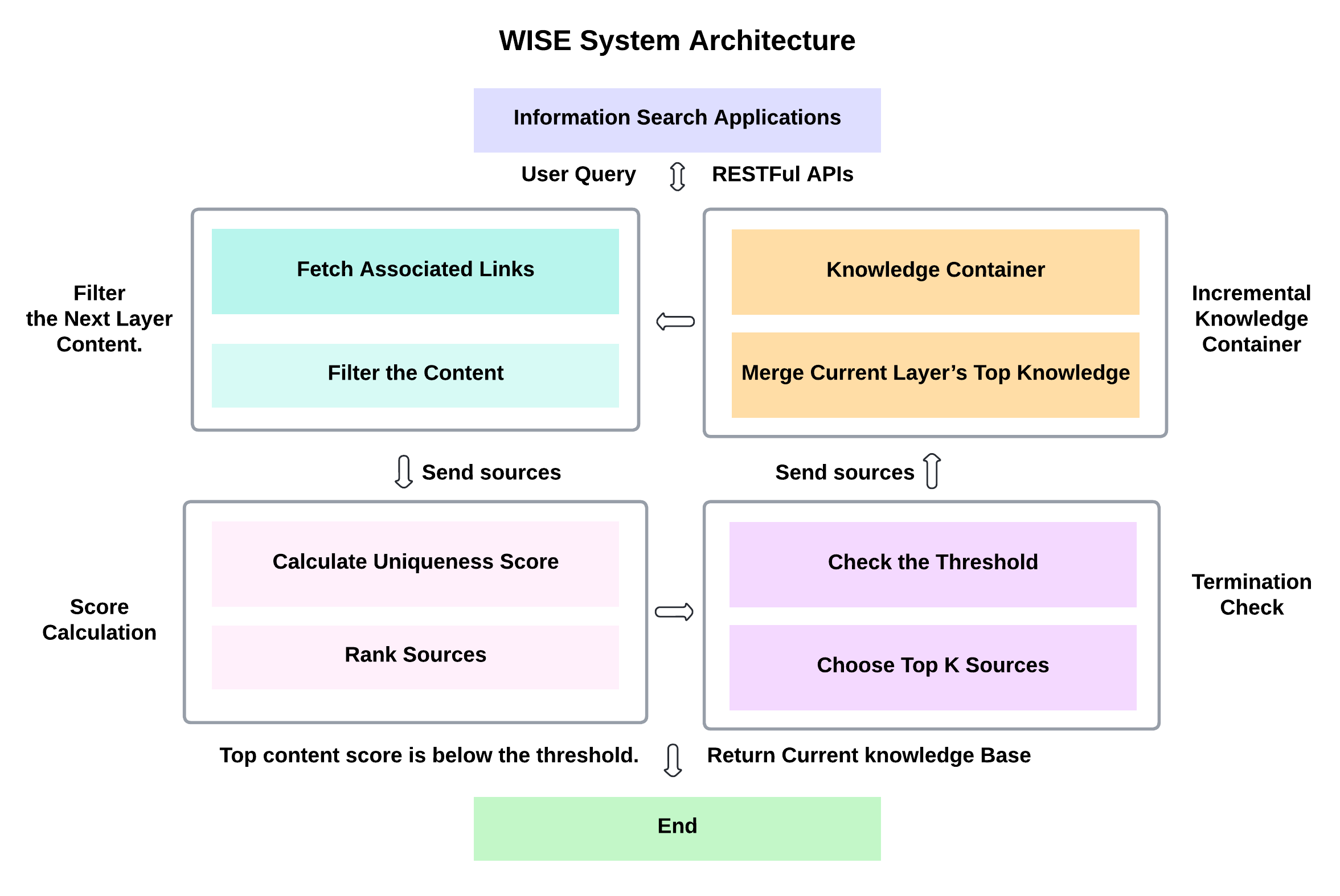}
 \caption{WISE System Architecture, showcasing its four main components: content filtering, score calculation, threshold checking, and knowledge consolidation.}
 \label{wise_system_architecture}
 \end{figure}

The WISE framework is designed to streamline the extraction and synthesis of knowledge from unstructured data sources through a structured and multi-layered approach. As illustrated in Figure~\ref{wise_system_architecture}, its architecture comprises four key stages that work in tandem to filter, score, rank, and consolidate information. Each stage contributes to transforming raw data into context-aware insights, enabling efficient knowledge discovery and refinement.

\begin{enumerate}
    \item \textbf{Content Filtering:} This stage employs query-specific extraction via LLM-driven contextual analysis, ensuring that only information relevant to the query is retained while noise, such as advertisements, is removed. This significantly reduces computational overhead in subsequent stages.
    \item \textbf{Score Calculation:} In this stage, the filtered content is evaluated for its unique contribution to the evolving knowledge container. Novel and relevant insights are prioritized, while redundant material is discarded.
    \item \textbf{Threshold Checking:} This component determines whether continued exploration of sources is justified. It acts as a termination criterion for the recursive process, halting when additional contributions fall below a defined threshold.
    \item \textbf{Knowledge Consolidation:} Extracted information is incrementally merged into a growing repository of domain-specific knowledge. This ensures that the final knowledge container is comprehensive, context-aware, and aligned with the user's query.
\end{enumerate}

WISE initiates its process with a user-provided query $q$ and an empty knowledge container $\mathcal{K}_0$. This container evolves iteratively as new insights are integrated. The initial set of sources $\mathcal{S}_0 = \{ s_1, s_2, \ldots, s_n \}$ is retrieved using a traditional similarity-based search function $\Phi(q)$, which identifies a collection of candidate sources relevant to the query. These sources serve as the root nodes for the tree-based search process. Formally:

\[
\mathcal{K}_0 = \emptyset, \quad \mathcal{S}_0 = \Phi(q)
\]

\subsection{Content Filtering}
\label{content_filtering}

Each source $s_i$ belonging to the set $\mathcal{S}_l$ at layer $l$ undergoes a query-specific refinement process. This process extracts content directly relevant to the query $q$. The filtering function $\Gamma$, which leverages the contextual understanding capabilities of an LLM, transforms the raw content $\mathcal{C}(s_i)$ of a source $s_i$ into a focused subset $\mathcal{F}(s_i)$:

\[
\mathcal{F}(s_i) = \Gamma(q, \mathcal{C}(s_i))
\]

By isolating only the most pertinent information, $\mathcal{F}(s_i)$ significantly reduces noise and irrelevant data. This streamlining enhances the efficiency of subsequent computational tasks and downstream processing stages. For simplicity, the result of the filtering operation for all sources at layer $l$ is denoted as:

\[
\mathcal{F}_l = \{\mathcal{F}(s_1), \mathcal{F}(s_2), \ldots, \mathcal{F}(s_n)\}_{l}
\]

Here, $\mathcal{F}_l$ represents the set of all query-specific filtered content derived at layer $l$ from the sources in $\mathcal{S}_l$.

\subsection{Score Calculation}
\label{score_calculation}

Following the filtering stage, WISE quantifies the unique contribution of each source to the knowledge container. Let $w_{\text{filtered}}(s_i)$ denote the number of words in the filtered content of each source $s_i$:

\[
w_{\text{filtered}}(s_i) = |\mathcal{F}(s_i)|
\]

where $|\mathcal{F}(s_i)|$ represents the cardinality (number of elements) of the set $\mathcal{F}(s_i)$. 

Next, we determine the number of words that overlap between the source's filtered content $\mathcal{F}(s_i)$ and the current knowledge container $\mathcal{K}_l$. Let $w_{\text{overlap}}(s_i, \mathcal{K}_l)$ denote this count:

\[
w_{\text{overlap}}(s_i, \mathcal{K}_l) = |\mathcal{F}(s_i) \cap \mathcal{K}_l|
\]

Here, $|\mathcal{F}(s_i) \cap \mathcal{K}_l|$ represents the cardinality of the intersection of the two sets. 

The unique knowledge contribution $\mathcal{K}(s_i)$ of source $s_i$ is then defined as the difference between the number of words in the filtered content and the number of overlapping words:

\[
\mathcal{K}(s_i) = w_{\text{filtered}}(s_i) - w_{\text{overlap}}(s_i, \mathcal{K}_l)
\]

This value, $\mathcal{K}(s_i)$, represents the number of new, unique words that source $s_i$ contributes to the knowledge container.

To normalize and evaluate the contribution of each source, we define the following metrics:

\textbf{1. Knowledge Density (Per-Word Normalization):} This metric normalizes the unique knowledge contribution by the size of the source (measured in the number of words in the filtered content). It is calculated as:

\[
\text{Knowledge Density}(s_i) = \frac{\mathcal{K}(s_i)}{w_{\text{filtered}}(s_i)}
\]

This ratio represents the proportion of unique words in the filtered content of source $s_i$.

\textbf{2. Knowledge Increase (Relative Growth):} This metric measures the relative contribution of the source to the existing knowledge container, expressed as a proportion of the current size of the knowledge container:

\[
\text{Knowledge Increase}(s_i) = \frac{\mathcal{K}(s_i)}{|\mathcal{K}_l|}
\]

Here, $|\mathcal{K}_l|$ denotes the cardinality of the knowledge container $\mathcal{K}_l$, representing the total number of words in the knowledge container at layer $l$.

To integrate the concepts of local efficiency (size of the source) and global contribution (size of the knowledge container), we define a unified scoring function $\Psi$ that employs log scaling to balance these factors:

\textbf{3. Combined Normalized Metric:}

\[
\text{Score}(s_i) = \Psi(\mathcal{F}(s_i), \mathcal{K}_l) = \frac{\mathcal{K}(s_i)}{\log(1 + w_{\text{filtered}}(s_i) + |\mathcal{K}_l|)}
\]

This combined metric prioritizes sources that offer unique and meaningful contributions, accounting for both the relative size of the source (in terms of the number of words in its filtered content) and its impact on the evolving knowledge container (in terms of the number of unique words it contributes).

\subsection{Threshold Checking and Pruning}
\label{threshold_checking_and_pruning}

WISE evaluates whether continued exploration will yield meaningful insights by comparing the highest score among the current sources, $\max_{s_i \in \mathcal{S}_l} \text{Score}(s_i)$, to a predefined threshold $\mathcal{T}$. If no source surpasses this threshold, the recursive process terminates:

\[
\max_{s_i \in \mathcal{S}_l} \text{Score}(s_i) < \mathcal{T} \implies \text{Terminate}
\]

If at least one source meets or exceeds the threshold, WISE selects the top $k$ sources based on their scores for further exploration:

\[
\mathcal{S}_{l+1} = \text{Top}_k(\mathcal{S}_l, \text{Score})
\]

This pruning step ensures that computational efforts are focused on sources most likely to enrich the knowledge container.

\subsection{Knowledge Container Construction}
\label{knowledge_container_construction}

The knowledge container $\mathcal{K}_l$ is updated by incorporating the filtered content of the chosen sources. This process further enhances the repository of query-relevant information. The update rule is defined as:

\[
\mathcal{K}_{l+1} = \Lambda(\mathcal{K}_l, \mathcal{S}_{l+1}) = \mathcal{K}_l \cup \bigcup_{s_i \in \mathcal{S}_{l+1}} \mathcal{F}(s_i)
\]

Here, $\Lambda$ represents an LLM-powered fusion function designed to merge new information from the selected sources $\mathcal{S}_{l+1}$ with the existing knowledge base $\mathcal{K}_l$. Upon termination of the recursive process, the final knowledge base $\mathcal{K}_f$ represents the aggregated and refined knowledge for the query $q$:

\[
\mathcal{K}_f = \mathcal{K}_l \quad \text{at termination}
\]

\subsection{Recursive Algorithm: WISE Framework}

Algorithm~\ref{alg:wise_recursive} outlines the recursive process for constructing a query-specific knowledge base. The subsequent layer's sources, $\mathcal{S}_{l+1}$, are obtained by analyzing the links embedded in the filtered content of the top $k$ sources from the current layer, $\text{Top}_k(\mathcal{S}_l, \text{Score})$. This ensures that the exploration focuses on paths that are both contextually relevant and computationally efficient.

\begin{algorithm}[h]
\caption{Recursive WISE Framework}
\label{alg:wise_recursive}
\raggedright 

\SetAlgoNlRelativeSize{-1} 
\SetAlgoVlined 
\SetKwInOut{Input}{\textbf{Input}}\SetKwInOut{Output}{\textbf{Output}}

\Input{Query $q$, Initial sources $\mathcal{S}_0$, Knowledge Container $\mathcal{K}_0$, Threshold $\mathcal{T}$}
\Output{Final knowledge base $\mathcal{K}_f$}

\SetKwFunction{WISE}{WISE}
\SetKwProg{Fn}{Function}{:}{}

\Fn{\WISE{$\mathcal{S}_l, \mathcal{K}_l, l$}}{
    \If{$\mathcal{S}_l = \emptyset$ \textbf{or} $\max_{s_i \in \mathcal{S}_l} \text{Score}(s_i) < \mathcal{T}$}{
        \Return $\mathcal{K}_l$ \tcp*{Terminate recursion if no significant sources remain or the set of sources is empty}
    }
    
    \ForEach{$s_i \in \mathcal{S}_l$}{
        $\mathcal{F}(s_i) \gets \Gamma(q, \mathcal{C}(s_i))$ \tcp*{Filter content for query $q$ using filtering function $\Gamma$}
        $\text{Score}(s_i) \gets \Psi(\mathcal{F}(s_i), \mathcal{K}_l)$ \tcp*{Calculate score for filtered content and knowledge base using scoring function $\Psi$}
    }
    
    $\mathcal{S}_{l+1} \gets \text{Top}_k(\mathcal{S}_l, \text{Score})$ \tcp*{Select top $k$ sources for the next layer based on their scores}
    $\mathcal{K}_{l+1} \gets \Lambda(\mathcal{K}_l, \mathcal{S}_{l+1})$ \tcp*{Update knowledge base by merging with selected sources using fusion function $\Lambda$}
    
    \Return \WISE{$\mathcal{S}_{l+1}, \mathcal{K}_{l+1}, l+1$} \tcp*{Recursive call to the next layer}
}

\BlankLine

\Return $\mathcal{K}_f \gets \text{WISE}(\mathcal{S}_0, \mathcal{K}_0, 0)$ \tcp*{Initialize recursion with initial sources, empty knowledge container, and layer 0}

\end{algorithm}

\section{Experiment}
\label{experiment}

The experimental setup and methodology used to evaluate WISE's ability to extract and synthesize knowledge from unstructured scientific data are detailed here. Our experiments focused on the query:

\begin{quote}
    \textit{Q: What is the comprehensive set of diseases and phenotypes that are linked to genetic variants within the HBB gene?}
\end{quote}

This query, centered on the \textit{HBB} gene, provided a rigorous test case for WISE's capabilities due to the domain's complexity and the interconnected nature of the information.

\subsection{Experimental Setup}
\label{experiment_setup}

The experiment started with an initial set of 24 sources related to the \textit{HBB} gene, obtained from the HGNC database \cite{hgnc4827}. Each source was meticulously classified into sections using structural elements, tags, and hyperlinks, extracted through a regular expression-based process. This resulted in a dataset enriched with metadata, including section identifiers and reference sources. We performed asynchronous content extraction from these sources, which served as the foundational nodes for WISE's progressive deepening process \cite{han1995mining}.

\subsection{Query-Specific Content Filtering}

WISE begins by employing the LLM-driven content filtering process detailed in Section \ref{content_filtering} to refine the raw content of each source based on its relevance to the user-specified query. For this experiment, focused on the query $Q$, the filtering function $\Gamma$ takes advantage of contextual understanding of the LLM to isolate pertinent information, eliminating extraneous content such as advertisements, unrelated sections, and general background information that does not directly address the specifics of the query.

Figure \ref{combined_visualization_of_layers} illustrates the significant impact of content filtering on data volume for selected sources, demonstrating the substantial reduction in content size achieved through this process. Notably, the UniProt \cite{uniprotP68871} entry for the \textit{HBB} gene, initially containing 8,249 words, was reduced to just 355 words after applying query-specific filtering. This exemplifies the prevalence of extraneous information even within highly regarded scientific resources. Across all sources, filtering reduced content size by an average of 80.14\%, with reductions as high as 96.12\% observed in some cases. This dramatic reduction highlights the effectiveness of LLM-driven filtering in isolating relevant content, thereby significantly reducing computational overhead and improving the efficiency and precision of downstream processing stages. By focusing on query-relevant information, WISE ensures that subsequent steps, such as score calculation and knowledge integration, operate on a refined and highly pertinent dataset.

\subsection{Score Calculation and Ranking}

To validate the effectiveness of our scoring mechanism in prioritizing query-specific relevance, we conducted experiments focusing on the query $Q$. As detailed in Section \ref{score_calculation}, WISE employs a dynamic, content-driven scoring approach that contrasts sharply with traditional ranking methods used by systems like Google and ChatGPT, which often rely on factors like source popularity or SEO (Search engine optimization) \cite{sharma2019brief} optimization.

\begin{figure}[h]
\centering
\includegraphics[width=1\linewidth]{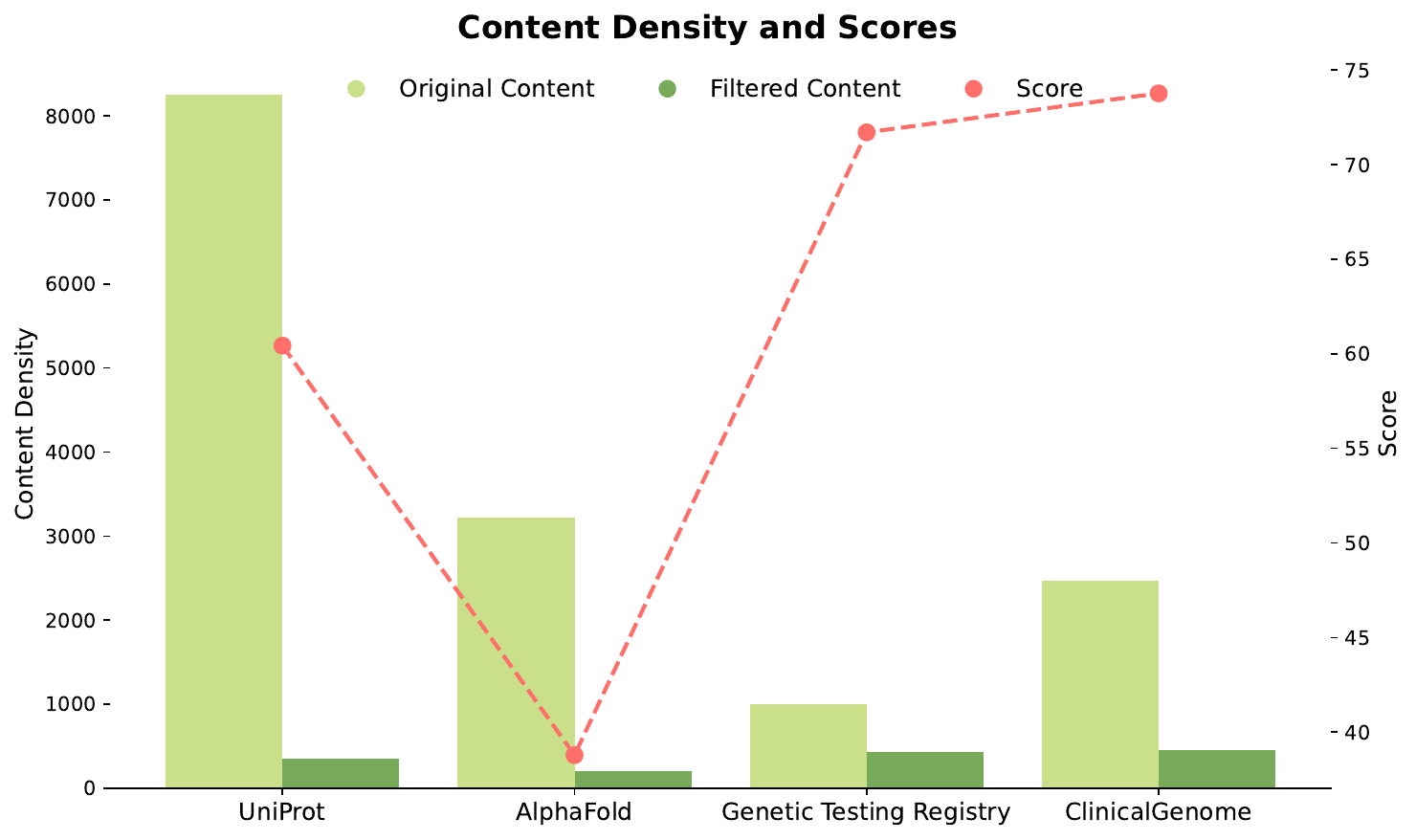}
\caption{WISE filtering and scoring in selected 4 sources, showing content density before and after filtering along with their scores. It demonstrates that content size alone does not determine a high score; the unique contribution must be significant relative to the existing knowledge base.}
\label{combined_visualization_of_layers}
\end{figure}

Our scoring mechanism calculates a combined normalized score for each source, integrating both local efficiency (source size) and global contribution (impact on the knowledge container). This ensures that sources offering substantive, query-specific insights are prioritized, regardless of their general popularity. Initial experiments, illustrated in Figure \ref{combined_visualization_of_layers}, demonstrate that widely recognized sources, such as UniProt \cite{uniprotP68871} or AlphaFold \cite{jumper2021highly}, did not always rank highest due to the presence of content unrelated to the specific query. This finding validates our design choice to prioritize content relevance over superficial attributes.

\subsection{Thresholding and Knowledge Container Construction}

WISE employs a threshold-based pruning mechanism, as detailed in Section \ref{threshold_checking_and_pruning}, to ensure efficient exploration of the knowledge space. This mechanism dynamically determines when to terminate the search process based on the diminishing returns observed in source scores. As the system progresses through successive layers, the knowledge container (described in Section \ref{knowledge_container_construction}) grows, leading to increased overlap between newly encountered sources and the existing knowledge. Consequently, the unique contribution of each new source tends to decrease.

In this experiment, a threshold value of 20 was empirically determined to effectively balance exploration and exploitation. When the highest score among the current sources falls below this threshold, or when no additional sources are available, WISE terminates the exploration process. This adaptive stopping criterion, mirroring the behavior of expert researchers, prevents the system from pursuing low-yield paths and conserves computational resources.

\begin{figure}[h]
\centering
\includegraphics[width=1\linewidth]{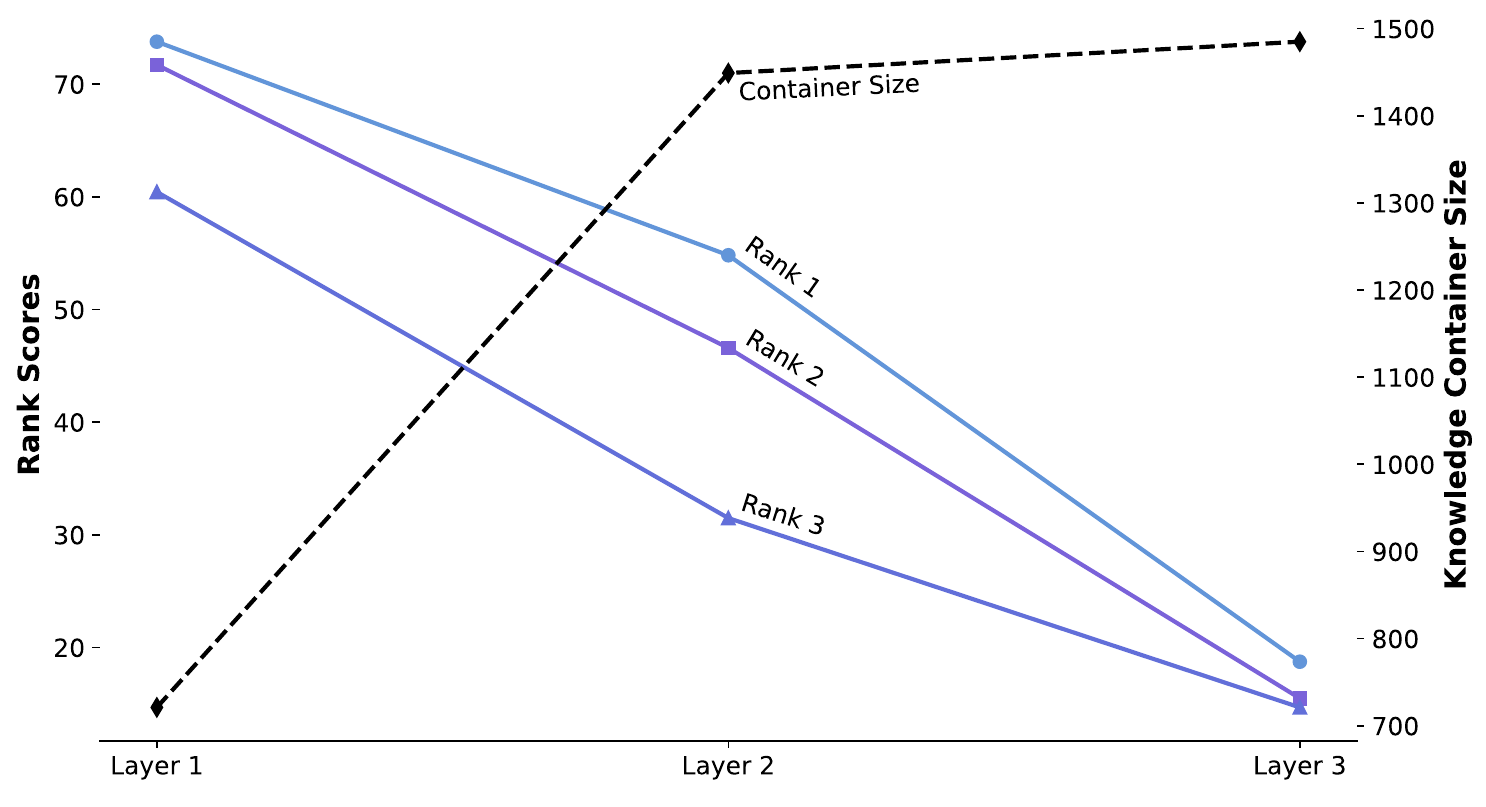}
\caption{Scores for Ranks 1, 2, and 3 across layers, highlighting their gradual decrease over time. This also demonstrates the growth of the knowledge container size with each layer, eventually plateauing as layers increase.}
\label{data_extraction_status_across_categories}
\end{figure}

Figure \ref{data_extraction_status_across_categories} illustrates this phenomenon, showing a consistent decrease in the scores of top-ranked sources across three successive layers. This trend validates the effectiveness of the threshold-based pruning strategy in identifying the point of diminishing returns. The knowledge container, constructed by integrating the filtered content from the top two sources at each layer, evolves into a rich and contextually relevant repository of information, progressively refined with each iteration. This iterative process ensures that the final knowledge container is both comprehensive and focused, containing the most valuable insights related to the query.

\section{Results}
\label{results}

A comprehensive analysis of WISE's performance, contrasting it with established methods across several critical dimensions, is presented here. Our evaluation, based on the experiments detailed in Section \ref{experiment}, focused on the query $Q$, designed to probe both the breadth and depth of knowledge extraction. To provide a robust and nuanced evaluation, we compared WISE against four baseline systems, each representing a different approach to information retrieval and synthesis: Pure ChatGPT \cite{app14177782, openai2024gpt4o} (a standard model, version GPT-4o accessed via the OpenAI API, relying solely on its pre-trained knowledge), ChatGPT with Search \cite{openai2023chatgptsearch, sun2023chatgpt} (a version of ChatGPT, version GPT-4o augmented with web search capabilities), Gemini \cite{team2023gemini} (Google's large language model, designed to integrate information from various sources), and traditional Google Search \cite{piasecki2018google, cilibrasi2007google} (the standard Google Search engine, considering the top 5 search results for analysis). For all baseline systems, default parameters were used, and the query $Q$ was directly input without any modifications.

\subsection{Evaluation Metrics}

Our analysis employs a combination of quantitative metrics, focusing on the relevance, uniqueness, and depth of the extracted information.

\subsubsection{ROUGE and BLEU}

\begin{figure*}[h]
    \centering
    \includegraphics[width=0.8\linewidth]{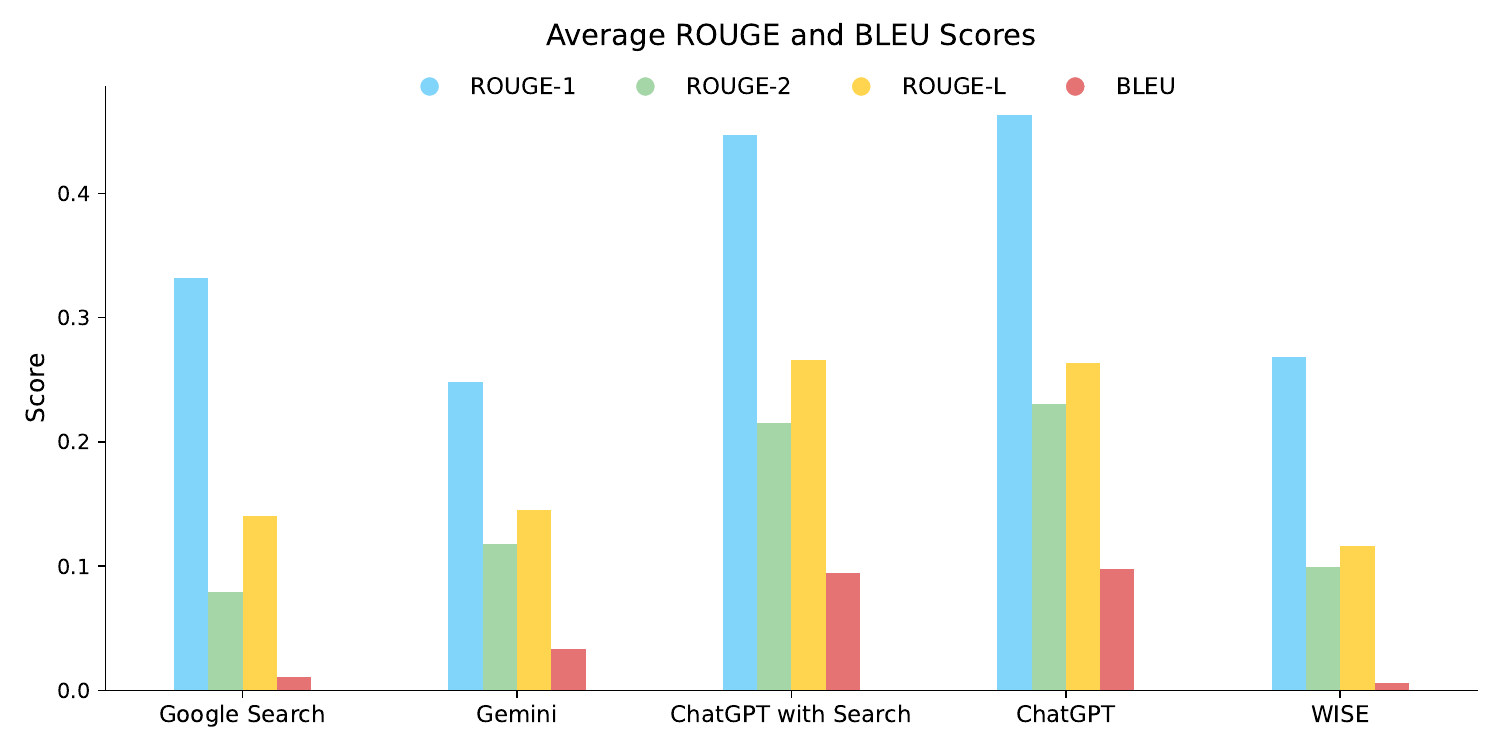}
    \caption{Average ROUGE and BLEU scores for each system when used as a reference, demonstrating that WISE's output is the most distinct and least repetitive.}
    \label{fig:average_rouge_bleu}
\end{figure*}

To assess the overlap and uniqueness of the information extracted by each system, we employed ROUGE \cite{lin2005recall} (Recall-Oriented Understudy for Gisting Evaluation) and BLEU \cite{papineni2002bleu} (Bilingual Evaluation Understudy) metrics, commonly used for evaluating machine-generated text. We adapted these metrics, calculating ROUGE-1, ROUGE-2, and ROUGE-L, along with BLEU, to compare each system's output against the others. Figure \ref{fig:average_rouge_bleu} presents the average ROUGE and BLEU scores for each system when used as a reference, revealing that WISE consistently exhibits the lowest scores, indicating that its generated content is more distinct and less repetitive compared to other approaches.

\subsubsection{Recall}

\begin{figure}[H]
\centering
\includegraphics[width=0.9\linewidth]{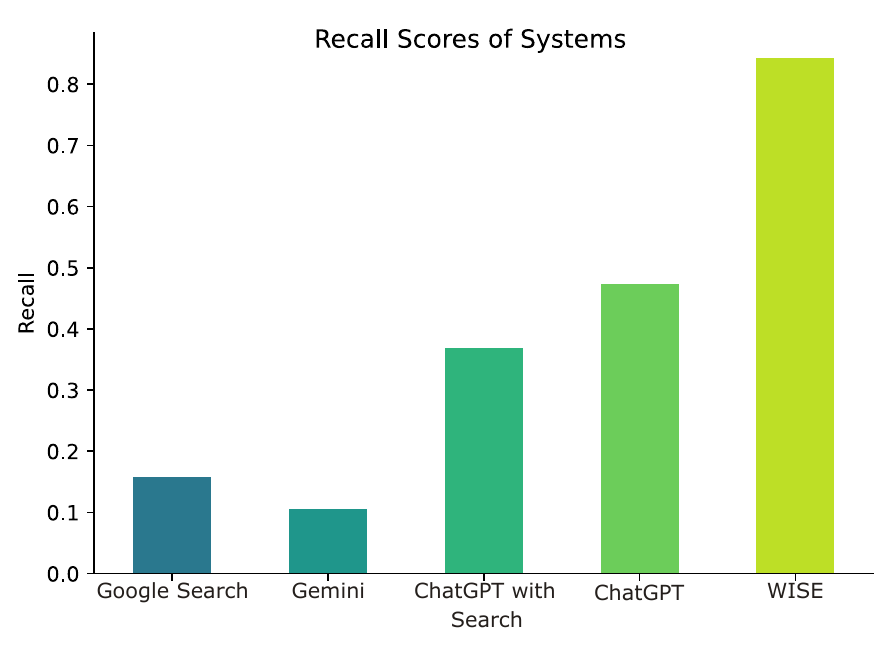}
\caption{Recall scores of each system, demonstrating that WISE identifies a greater proportion of diseases from the combined output.}
\label{fig:recall_scores}
\end{figure}

To evaluate the comprehensiveness of each system, we calculated their recall based on a combined output created by taking the union of all unique diseases identified by any of the five systems. This combined output, representing a comprehensive collection of potentially relevant diseases, is detailed in Table \ref{tab:pseudo_ground_truth}. WISE achieved a recall of 0.842, significantly outperforming the baseline systems, as shown in Figure \ref{fig:recall_scores}. In contrast, ChatGPT achieved a recall of 0.474, ChatGPT with Search achieved a recall of 0.368, while Google Search Gemini and traditional Google Search scored considerably lower at 0.105 and 0.158 respectively. This emphasizes WISE's superior ability to identify a greater proportion of potentially relevant diseases.

\begin{table*}[h]
    \centering
    \caption{Disease Identification Across Systems}
    \begin{tabular}{lccccc}
        \hline
        \textbf{Disease} & \textbf{Normal Google Search} & \textbf{Google Search Gemini} & \textbf{ChatGPT with Search} & \textbf{ChatGPT} & \textbf{WISE} \\
        \hline
        Hemoglobin SC & \xmark & \xmark & \cmark & \cmark & \cmark \\
        Hemoglobin O & \xmark & \xmark & \xmark & \cmark & \xmark \\
         Hemoglobin S/\ensuremath{\beta}-Thalassemia & \xmark & \xmark & \cmark & \cmark & \cmark \\
        Hemoglobin S Oman & \xmark & \xmark & \xmark & \xmark & \cmark \\
        Malaria & \xmark & \xmark & \xmark & \xmark & \cmark \\
        Sickle Cell Disease & \cmark & \cmark & \cmark & \cmark & \cmark \\
        Hispanic Gamma-Delta-\ensuremath{\beta} Thalassemia & \xmark & \xmark & \xmark & \xmark & \cmark \\
        \ensuremath{\beta}-Type Methemoglobinemia & \xmark & \xmark & \xmark & \xmark & \cmark \\
        Dominant \ensuremath{\beta}-Thalassemia & \xmark & \xmark & \xmark & \xmark & \cmark \\
        Hemoglobinopathies & \cmark & \xmark & \xmark & \xmark & \xmark \\
        Heinz Body Anemia & \xmark & \xmark & \xmark & \xmark & \cmark \\
        Hemoglobin C & \xmark & \xmark & \cmark & \cmark & \cmark \\
        Hemoglobin M & \xmark & \xmark & \xmark & \xmark & \cmark \\
        Hemoglobin D & \xmark & \xmark & \xmark & \cmark & \xmark \\
         Familial Erythrocytosis 6 & \xmark & \xmark & \xmark & \xmark & \cmark \\
        Hemoglobin S Antilles & \xmark & \xmark & \xmark & \xmark & \cmark \\
        Hemoglobin E & \xmark & \xmark & \cmark & \cmark & \cmark \\
       Hereditary Persistence of Fetal Hemoglobin & \xmark & \xmark & \cmark & \cmark & \cmark \\
       \ensuremath{\beta}-Thalassemia & \cmark & \cmark & \cmark & \cmark & \cmark \\
        \hline
    \end{tabular}
    \label{tab:pseudo_ground_truth}
\end{table*}

\subsubsection{Level-Based Analysis}

\begin{figure*}[h]
    \centering
    \includegraphics[width=0.8\linewidth]{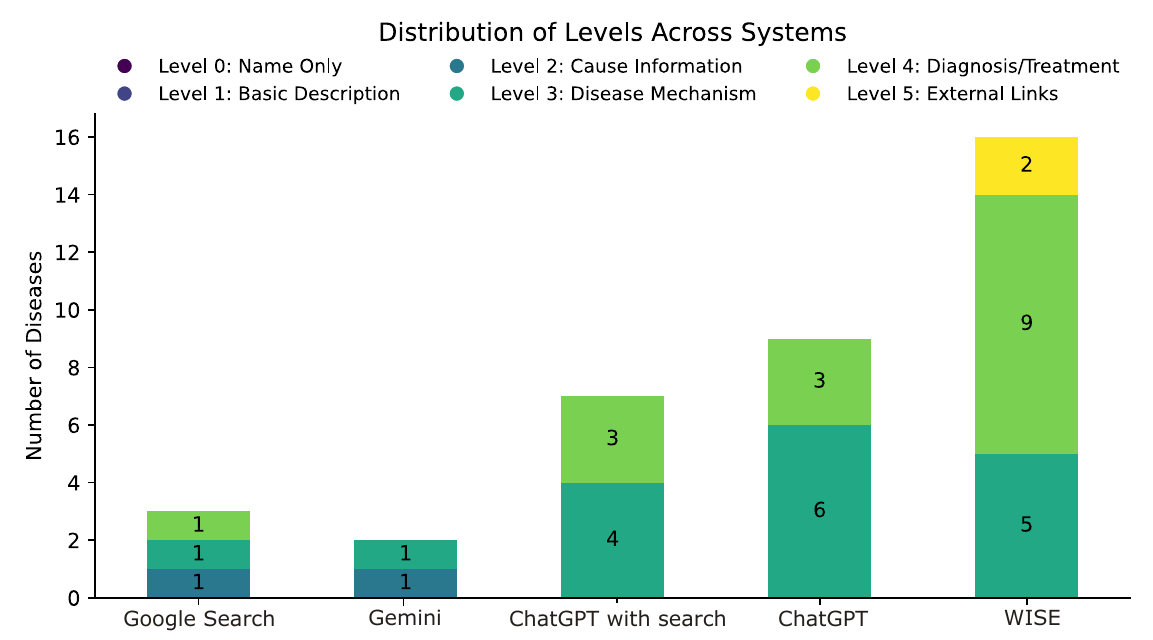}
    \caption{Average level of detail for each system, demonstrating WISE's superior ability to provide in-depth and comprehensive information about identified diseases.}
    \label{fig:average_level_scores}
\end{figure*}

To further analyze the depth of information provided by each system, and to establish a metric that can generally be used to assess the richness of content, we employed a level-based analysis. In this approach, each identified disease was manually assigned a level from 0 to 5 based on the following criteria:
\begin{itemize}
    \item \textbf{Level 0:} Disease name only.
    \item \textbf{Level 1:} Basic description of the disease.
    \item \textbf{Level 2:} Information about the cause of the disease.
    \item \textbf{Level 3:} Details about the disease's mechanism.
    \item \textbf{Level 4:} Information about diagnosis, treatment, or prognosis.
    \item \textbf{Level 5:} Links to external resources or discussion of ongoing research.
\end{itemize}
This level-based approach, while applied to evaluate WISE and baselines in this experiment, also offers a generalizable method for assessing the richness of information provided by any system. Figure \ref{fig:average_level_scores} presents the average level of detail for each system, showing that WISE achieved the highest score (3.81), significantly surpassing the other systems. This result demonstrates that WISE provides more in-depth and comprehensive information about the identified diseases compared to the baselines.

The convergence of these metrics paints a compelling picture of WISE's strengths. It is not only capable of identifying a wider range of diseases and phenotypes linked to the \textit{HBB} gene (demonstrated by its high recall) but also of providing richer, more unique, and contextually relevant information (shown through the ROUGE, BLEU, and level-based analyses). The superior performance of WISE across all these metrics highlights its potential as a transformative system for information retrieval in complex domains.

These findings can be attributed to WISE’s unique design, particularly its dynamic scoring mechanism, tree-based architecture, and LLM-powered content filtering. The dynamic scoring prioritizes content relevance over superficial attributes, ensuring that the most valuable sources are identified. The tree-based architecture allows for efficient exploration of the knowledge space, while the LLM-driven filtering ensures that only pertinent information is processed.

\section{Applications}
\label{applications}

The adaptability of WISE extends far beyond the specific use case we have explored thus far, underscoring its potential to transform knowledge synthesis across diverse domains. In this section, we illustrate WISE's versatility by exploring its prospective applications, demonstrating how its unique capabilities can address existing challenges and accelerate progress in various fields.

\subsection{Drug Discovery: Unveiling Novel Therapeutic Pathways}

The process of drug discovery often hinges on unraveling the intricate relationships between genes, diseases, and potential therapeutic targets. WISE offers a transformative approach to this challenge by providing an efficient and comprehensive means of identifying these complex associations, surpassing the limitations of manual methods and existing systems. Consider the following query:

\begin{quote}
	Q: \textit{What diseases are associated with \textit{C16orf82}, and are there any existing drugs targeting these conditions?}
\end{quote}

This query, while seemingly simple, requires navigating a complex web of genetic and pharmacological information. WISE can reveal novel connections within the existing literature, identifying not only diseases sharing genetic origins or structural similarities in proteins (including overlapping reading frames, ORFs) but also highlighting previously unlinked diseases that may share common pathways or molecular interactions. These insights provide valuable leads for drug repurposing and the development of novel therapeutic strategies, effectively accelerating the drug discovery process by integrating these discoveries with relevant drug information, thus delivering precise and actionable intelligence for pharmaceutical development.

\subsection{Material Structure Analysis: Accelerating Inverse Design}

The field of materials science, particularly the domain of inverse material design, is often constrained by time-intensive and inefficient processes for identifying suitable material structures that meet specific requirements. WISE offers a streamlined approach to this challenge. For instance, consider this query:

\begin{quote}
	Q: \textit{What are the most suitable material structures for achieving high thermal conductivity and mechanical strength in lightweight applications?}
\end{quote}

By directly linking structural properties to specific application requirements, WISE enables researchers to explore material structure databases with unprecedented speed and efficiency. This reduces the manual effort and time required for material selection, allowing researchers to focus on designing solutions that precisely meet their objectives, rather than spending excessive time on information gathering. By reducing reliance on inefficient, time-intensive methods, WISE serves as a powerful tool to accelerate the field of materials science.

\subsection{Social Issue Analysis: Illuminating Complex Societal Challenges}

WISE is equally applicable to addressing complex social issues, where the analysis of vast amounts of unstructured data is critical. Social scientists and policymakers grapple with a range of complex challenges, requiring the integration of data from diverse sources to identify patterns and develop effective interventions. For example:

\begin{quote}
	Q: \textit{What factors are contributing to the rising cancer rates in [specific location]?}
\end{quote}

This query, designed to highlight the social, environmental, and economic challenges that drive increases in rates of cancer, demands the integration of multiple viewpoints, datasets, and research findings. WISE can synthesize information from diverse sources to identify complex patterns, including increased exposure to environmental toxins, socio-economic inequalities, or shortcomings in public health policy. By highlighting key contributing factors, WISE offers researchers and policymakers critical data points and insight that empowers them to develop data-driven hypotheses and implement more targeted interventions. The ability of WISE to generate comparative examples from similar regions further allows for a deeper, more nuanced understanding of the issue at hand.

The examples above underscore WISE's flexibility and adaptability, demonstrating its applicability beyond specific domains. Its ability to process complex queries and synthesize domain-specific knowledge makes it a valuable asset across a broad range of fields, from medical research to materials science and social issue analysis. Over time, WISE's workflow has the potential to further accelerate progress in areas like cancer research, environmental studies, and many others, by providing a more efficient and reliable approach to obtaining detailed and context-aware information. These diverse applications highlight WISE's potential to enable deeper insights and foster innovation across a wide spectrum of scientific and societal disciplines.

\section{Future Work}
\label{future_work}

While WISE has demonstrated significant capabilities in our experiments, its journey is far from complete. We are actively exploring several promising avenues for further development, poised to enhance the system's robustness, efficiency, and applicability. The following represent key directions for future research, although these improvements are not yet incorporated into the current implementation and remain outside the scope of this paper.

\subsection{Knowledge Graph Integration: Unlocking Deeper Relational Insights}

The integration of knowledge graphs represents a transformative opportunity to amplify WISE's ability to reason about complex relationships within scientific data. Knowledge graphs, which represent information as interconnected nodes and edges, offer a structured approach for preserving and reasoning about intricate interdependencies. Such an approach transcends the limitations of purely text-based analysis, enabling WISE to identify connections between seemingly disparate entities and uncover subtle yet significant articulation points that are often missed by traditional methods.

By incorporating knowledge graphs, WISE can maintain a dynamic understanding of the relationships between entities, thereby eliminating the need for repeated, LLM-driven knowledge unions. Instead, newly extracted information can be directly appended to the appropriate nodes and edges in the graph, ensuring continuity, efficiency, and preventing information loss. Preliminary experiments, for example, have demonstrated that representing the UniProt entry for the \textit{HBB} gene as a knowledge graph with 56 nodes and 55 edges effectively captures its content with reduced complexity compared to raw text. Further filtering this knowledge graph with a query related to \textit{HBB}-specific diseases resulted in a focused subgraph with 11 nodes and 16 edges, while maintaining key interconnected causes, like shared hormonal pathways across multiple diseases.

Furthermore, knowledge graphs facilitate intuitive visualizations, enhancing user understanding and interpretation. Their structured nature supports advanced reasoning capabilities, enabling WISE to achieve deeper insights through graph-based matching techniques, outperforming traditional content comparison approaches. This integration promises a more powerful and efficient means of knowledge discovery.

\subsection{Enhanced Query Engagement: Steering Towards Precise Intent}

WISE currently relies on user-provided queries, future iterations will focus on enhancing query engagement to steer the system towards a more precise understanding of user intent. Although our similarity-based searches have proven effective so far, there is a clear potential to amplify WISE’s performance through a more iterative and user-involved query process. Future iterations of WISE will incorporate mechanisms for better understanding user intent through supplementary information gathering. For example, users could be prompted to provide additional context or goals for their search, enabling more precise and targeted results.

Moreover, the system could implement automatic query enhancement, leveraging prior searches and literature data to refine user input iteratively. This process may also include layers of semantic understanding, improved similarity measures, and propose augmented queries for user approval. These advancements would significantly reduce the burden on less-experienced users, simplifying complex search tasks while maintaining the high standards of precision that WISE offers. These improvements could also help the system identify if the query is underdefined or if there is some implicit constraints in the query.


\section{Related Work}

Prior research in information extraction has significantly advanced from manual curation and feature-engineered machine learning models to more sophisticated LLM-based approaches. Systems such as GIX \cite{10.1371/journal.pone.0303231} effectively leverage large language models to automate gene interaction extraction, outperforming earlier methods on benchmark datasets. Similarly, tree-structured neural architectures have improved the identification of protein-protein interactions \cite{8665584}, and comprehensive reviews highlight the rise of neural network-based classifiers in biomedical relation extraction \cite{10.1016/j.jbi.2019.103294}. Domain-adapted models, such as BioBERT \cite{lee2020biobert}, have demonstrated notable gains in precision for tasks including named entity recognition and relation extraction, while further explorations have extended the scope from binary to complex biomedical relations \cite{https://doi.org/10.1155/2014/298473}. Beyond the biomedical domain, research has shown that integrating pre-trained models with domain-specific corpora and graph-based reasoning can uncover intricate patterns, enabling richer insights and improved retrieval accuracy \cite{sarmah2024hybridrag, Buehler_2024, Zuluaga_Robledo_Arbelaez-Echeverri_Osorio-Zuluaga_Duque-Méndez_2022, 10.1007/s10994-021-06068-6}.

Despite these advancements, existing approaches often struggle to dynamically refine their focus or determine when further exploration yields diminishing returns. Retrieval-Augmented Generation techniques \cite{yu2024multisourceknowledgepruningretrievalaugmented} and workflow orchestration strategies \cite{fan2024workflowllmenhancingworkfloworchestration} have attempted to address these challenges by pruning sources and enhancing verification, yet they rarely employ hierarchical, query-driven frameworks that integrate filtering, scoring, and adaptive stopping criteria. Other efforts have focused on bridging the gap between unstructured and structured data \cite{10.1007/s11837-021-04902-9, xie2024bytesciencebridgingunstructuredscientific} but have not fully embraced iterative, tree-based methodologies for source selection and knowledge consolidation. In this context, WISE advances the state of the art by combining robust filtering, dynamic ranking, and a scalable tree-inspired workflow, thereby ensuring that only the most valuable, context-relevant information is retained for efficient, high-quality knowledge extraction.

\section{Discussion}

\begin{figure}[h]
	\centering
	\includegraphics[width=1\linewidth]{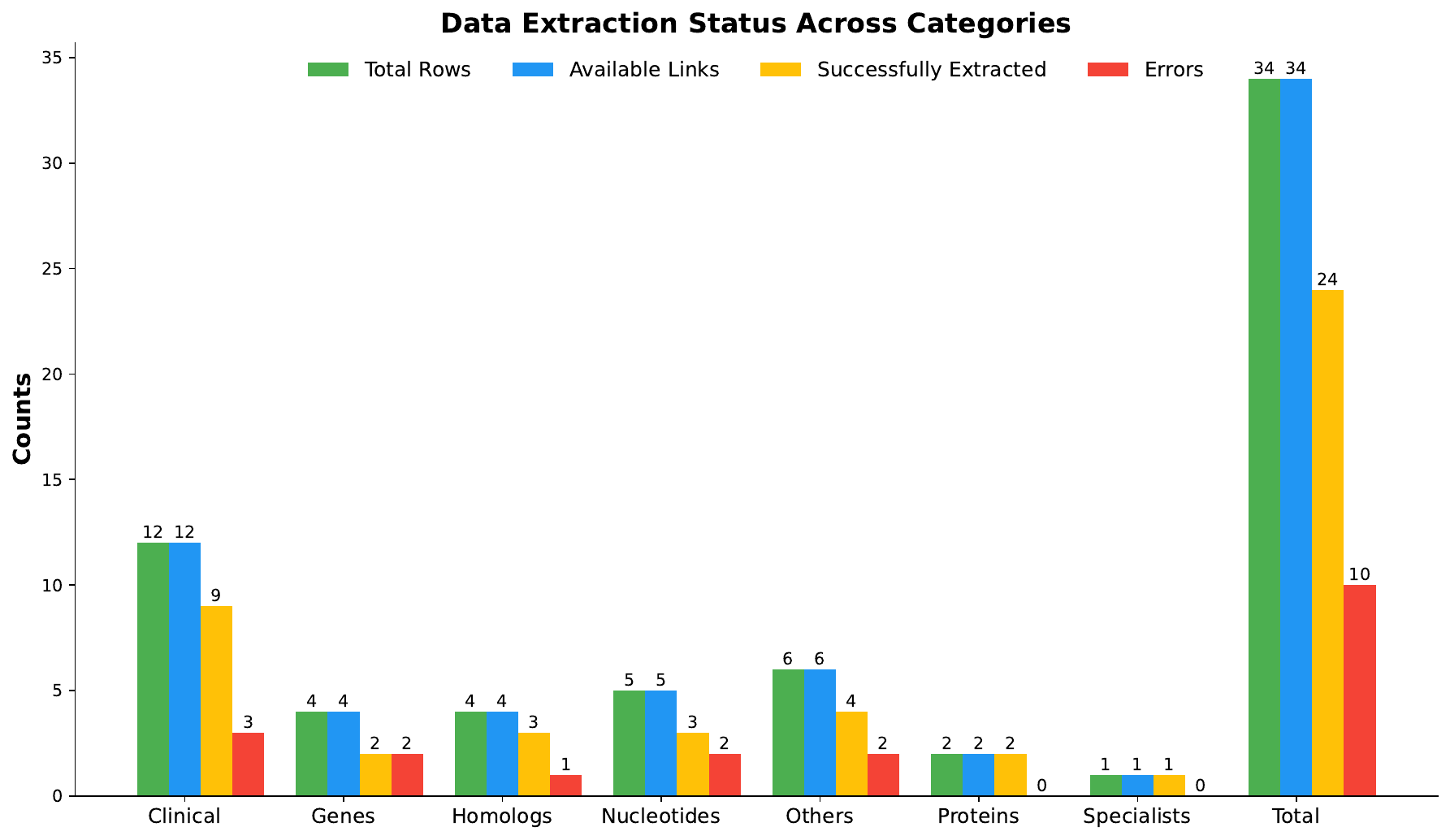}
	\caption{The number of sources that restrict data extraction by implementing blocks on automated processes.}
	\label{fig:data_extraction_status}
\end{figure}

The development and application of WISE highlight its transformative potential in synthesizing knowledge for complex queries, yet certain challenges and limitations merit discussion. One of the primary obstacles encountered during the experiments was restricted access to some data sources. Out of 34 initial sources, WISE successfully extracted content from 24, while the remaining 10 sources were inaccessible. Figure~\ref{fig:data_extraction_status} illustrates this disparity, emphasizing the growing trend among web platforms to limit data extraction. This failure to extract content stemmed primarily from two key factors: the increasing prevalence of paywalls, which create direct economic barriers to access, and security measures like bot detection, which are often deployed to protect intellectual property, user privacy, and prevent the unauthorized use of data for LLM training and content analysis. While these restrictions serve important purposes, such as protecting content creators and user data, they also hinder innovations like WISE, which focus on advancing non-commercial, academic, and research applications. Overcoming these challenges may require collaboration with data providers, the development of ethical and compliant retrieval techniques.

Another important consideration is the comprehensiveness of WISE’s outputs. By design, WISE delivers detailed, authoritative responses that include exhaustive references, disease sub-variations, and contextual information. While this level of detail is highly valuable for academic and clinical professionals, it may overwhelm non-specialist users who require more concise and simplified information. The extensive details and length of the responses may be daunting, highlighting the need for customizable output formats tailored to different audiences. Features such as adjustable levels of detail or user-specific summaries could enhance WISE’s accessibility and usability across a broader range of users.

A related challenge lies in word weighting during the synthesis of information. In WISE’s current implementation, more frequent words like “and” or “the” are deprioritized based on term frequency (TF), ensuring that the system focuses on content-specific terms. However, the exact contextual relationships between terms—critical for disambiguating similar entities—could benefit from improvements. Techniques like TF-IDF (Term Frequency-Inverse Document Frequency) are currently in consideration, as they would assign higher weight to less frequent but more meaningful terms. Additionally, as discussed in Section~\ref{future_work}, the integration of a knowledge graph offers a promising solution to this challenge. By representing relationships explicitly through nodes and edges, a knowledge graph would inherently prioritize meaningful connections, eliminating reliance on textual frequency metrics.

Despite these challenges, WISE represents a significant advancement in information retrieval and synthesis, setting a new standard for addressing complex queries. Its ability to dynamically filter, rank, and construct comprehensive knowledge containers demonstrates its transformative potential across diverse domains. The innovative architecture of WISE effectively bridges critical gaps in existing systems, offering a robust tool for academic, clinical, and interdisciplinary applications.

\section{Conclusion}

WISE presents a novel and effective approach to navigating the complexities of scientific information retrieval. By integrating LLM-driven filtering, dynamic ranking, and adaptive stopping criteria within a tree-based framework, WISE empowers researchers to efficiently and accurately extract and synthesize knowledge from vast and heterogeneous data sources. Our experiments on gene-disease association queries demonstrated WISE's superior performance compared to baseline methods, showcasing its ability to uncover a broader range of relevant information, including rare conditions and nuanced connections often overlooked by traditional search engines and basic LLM implementations. This enhanced precision and comprehensiveness, achieved through a content-driven, progressive deepening approach, offers significant potential for accelerating scientific discovery across diverse domains.

The development of WISE represents a substantial step forward in the pursuit of intelligent knowledge discovery. Its human-inspired methodology, mimicking the systematic approach of expert researchers, allows for a balanced exploration of information, prioritizing high-value insights while effectively managing computational resources. The framework's adaptability and scalability, demonstrated through its application to diverse research domains, further suggest its potential as a generalizable solution for complex information landscapes. We believe that WISE offers a valuable tool for researchers seeking to unlock the full potential of the ever-expanding universe of scientific knowledge, paving the way for more efficient and impactful research endeavors. By addressing the limitations of traditional search engines and general-purpose LLMs, WISE provides a robust and scalable solution for extracting and synthesizing knowledge, ultimately contributing to more informed and accelerated scientific progress.




\section*{Acknowledgment}

This Research was supported in part by a National Institutes of Health IDeA grant P20GM103408, a National Science Foundation CSSI grant OAC 2410668, and a US Department of Energy grant DE-0011014.

\bibliographystyle{IEEEtran}
\bibliography{sample-base}

\begin{thebibliography}{10}
\providecommand{\url}[1]{#1}
\csname url@samestyle\endcsname
\providecommand{\newblock}{\relax}
\providecommand{\bibinfo}[2]{#2}
\providecommand{\BIBentrySTDinterwordspacing}{\spaceskip=0pt\relax}
\providecommand{\BIBentryALTinterwordstretchfactor}{4}
\providecommand{\BIBentryALTinterwordspacing}{\spaceskip=\fontdimen2\font plus
\BIBentryALTinterwordstretchfactor\fontdimen3\font minus \fontdimen4\font\relax}
\providecommand{\BIBforeignlanguage}[2]{{%
\expandafter\ifx\csname l@#1\endcsname\relax
\typeout{** WARNING: IEEEtran.bst: No hyphenation pattern has been}%
\typeout{** loaded for the language `#1'. Using the pattern for}%
\typeout{** the default language instead.}%
\else
\language=\csname l@#1\endcsname
\fi
#2}}
\providecommand{\BIBdecl}{\relax}
\BIBdecl

\bibitem{10.1371/journal.pdig.0000347}
\BIBentryALTinterwordspacing
J.~Sedlakova, P.~Daniore, A.~Horn~Wintsch, M.~Wolf, M.~Stanikic, C.~Haag, C.~Sieber, G.~Schneider, K.~Staub, D.~Alois~Ettlin, O.~Grübner, F.~Rinaldi, V.~von Wyl, and for the University~of Zurich Digital Society Initiative (UZH-DSI) Health~Community, ``Challenges and best practices for digital unstructured data enrichment in health research: A systematic narrative review,'' \emph{PLOS Digital Health}, vol.~2, no.~10, pp. 1--22, 10 2023. [Online]. Available: \url{https://doi.org/10.1371/journal.pdig.0000347}
\BIBentrySTDinterwordspacing

\bibitem{doi:10.4137/BII.S31559}
\BIBentryALTinterwordspacing
J.~Luo, M.~Wu, D.~Gopukumar, and Y.~Zhao, ``Big data application in biomedical research and health care: A literature review,'' \emph{Biomedical Informatics Insights}, vol.~8, p. BII.S31559, 2016, pMID: 26843812. [Online]. Available: \url{https://doi.org/10.4137/BII.S31559}
\BIBentrySTDinterwordspacing

\bibitem{zhai2024large}
C.~Zhai, ``Large language models and future of information retrieval: Opportunities and challenges,'' in \emph{Proceedings of the 47th International ACM SIGIR Conference on Research and Development in Information Retrieval}, 2024, pp. 481--490.

\bibitem{10.1145/3626772.3657733}
\BIBentryALTinterwordspacing
A.~Salemi and H.~Zamani, ``Towards a search engine for machines: Unified ranking for multiple retrieval-augmented large language models,'' in \emph{Proceedings of the 47th International ACM SIGIR Conference on Research and Development in Information Retrieval}, ser. SIGIR '24.\hskip 1em plus 0.5em minus 0.4em\relax New York, NY, USA: Association for Computing Machinery, 2024, p. 741–751. [Online]. Available: \url{https://doi.org/10.1145/3626772.3657733}
\BIBentrySTDinterwordspacing

\bibitem{ziems2023large}
N.~Ziems, W.~Yu, Z.~Zhang, and M.~Jiang, ``Large language models are built-in autoregressive search engines,'' \emph{arXiv preprint arXiv:2305.09612}, 2023.

\bibitem{hgnc4827}
\BIBentryALTinterwordspacing
{HUGO Gene Nomenclature Committee (HGNC)}, ``{HBB Gene - Gene Symbol Report},'' 2024, accessed: 2024-12-11. [Online]. Available: \url{https://www.genenames.org/data/gene-symbol-report/\#!/hgnc_id/HGNC:4827}
\BIBentrySTDinterwordspacing

\bibitem{clinicalgenome_hgnc4827}
\BIBentryALTinterwordspacing
{Clinical Genome Resource (ClinGen)}, ``{HBB Gene - Clinical Genome Knowledge Base},'' 2024, accessed: 2024-12-11. [Online]. Available: \url{https://search.clinicalgenome.org/kb/genes/HGNC:4827}
\BIBentrySTDinterwordspacing

\bibitem{ncbi_nbk1435}
\BIBentryALTinterwordspacing
{National Center for Biotechnology Information (NCBI)}, \emph{{Sickle Cell Anemia - NCBI Bookshelf}}, 2024, accessed: 2024-12-11. [Online]. Available: \url{https://www.ncbi.nlm.nih.gov/books/NBK1435/}
\BIBentrySTDinterwordspacing

\bibitem{garcia2024reviewscientificknowledgeextraction}
\BIBentryALTinterwordspacing
G.~L. Garcia, J.~R.~R. Manesco, P.~H. Paiola, L.~Miranda, M.~P. de~Salvo, and J.~P. Papa, ``A review on scientific knowledge extraction using large language models in biomedical sciences,'' 2024. [Online]. Available: \url{https://arxiv.org/abs/2412.03531}
\BIBentrySTDinterwordspacing

\bibitem{systems11070351}
\BIBentryALTinterwordspacing
A.~Alshami, M.~Elsayed, E.~Ali, A.~E.~E. Eltoukhy, and T.~Zayed, ``Harnessing the power of chatgpt for automating systematic review process: Methodology, case study, limitations, and future directions,'' \emph{Systems}, vol.~11, no.~7, 2023. [Online]. Available: \url{https://www.mdpi.com/2079-8954/11/7/351}
\BIBentrySTDinterwordspacing

\bibitem{10020725}
\BIBentryALTinterwordspacing
S.~Saxena, R.~Sangani, S.~Prasad, S.~Kumar, M.~Athale, R.~Awhad, and V.~Vaddina, ``{ Large-Scale Knowledge Synthesis and Complex Information Retrieval from Biomedical Documents },'' in \emph{2022 IEEE International Conference on Big Data (Big Data)}.\hskip 1em plus 0.5em minus 0.4em\relax Los Alamitos, CA, USA: IEEE Computer Society, Dec. 2022, pp. 2364--2369. [Online]. Available: \url{https://doi.ieeecomputersociety.org/10.1109/BigData55660.2022.10020725}
\BIBentrySTDinterwordspacing

\bibitem{sneyd-stevenson-2019-modelling}
\BIBentryALTinterwordspacing
A.~Sneyd and M.~Stevenson, ``Modelling stopping criteria for search results using {P}oisson processes,'' in \emph{Proceedings of the 2019 Conference on Empirical Methods in Natural Language Processing and the 9th International Joint Conference on Natural Language Processing (EMNLP-IJCNLP)}, K.~Inui, J.~Jiang, V.~Ng, and X.~Wan, Eds.\hskip 1em plus 0.5em minus 0.4em\relax Hong Kong, China: Association for Computational Linguistics, Nov. 2019, pp. 3484--3489. [Online]. Available: \url{https://aclanthology.org/D19-1351/}
\BIBentrySTDinterwordspacing

\bibitem{10.1007/978-3-031-75434-0_23}
\BIBentryALTinterwordspacing
M.~Parmentier and A.~Legay, ``Adaptive stopping algorithms based on concentration inequalities,'' in \emph{Bridging the Gap Between AI and Reality: Second International Conference, AISoLA 2024, Crete, Greece, October 30 – November 3, 2024, Proceedings}.\hskip 1em plus 0.5em minus 0.4em\relax Berlin, Heidelberg: Springer-Verlag, 2025, p. 336–353. [Online]. Available: \url{https://doi.org/10.1007/978-3-031-75434-0_23}
\BIBentrySTDinterwordspacing

\bibitem{vaswani2023attentionneed}
\BIBentryALTinterwordspacing
A.~Vaswani, N.~Shazeer, N.~Parmar, J.~Uszkoreit, L.~Jones, A.~N. Gomez, L.~Kaiser, and I.~Polosukhin, ``Attention is all you need,'' 2023. [Online]. Available: \url{https://arxiv.org/abs/1706.03762}
\BIBentrySTDinterwordspacing

\bibitem{openai2024gpt4o}
\BIBentryALTinterwordspacing
OpenAI, ``Gpt-4o,'' 2024, the GPT-4o model supports a context window of up to 128,000 tokens. [Online]. Available: \url{https://platform.openai.com/docs/models/gpt-4-and-gpt-4-turbo}
\BIBentrySTDinterwordspacing

\bibitem{uniprotP68871}
\BIBentryALTinterwordspacing
U.~Consortium, ``Hemoglobin subunit beta,'' 2024, accessed: 2024-12-11. [Online]. Available: \url{https://www.uniprot.org/uniprotkb/P68871}
\BIBentrySTDinterwordspacing

\bibitem{han1995mining}
J.~Han, ``Mining knowledge at multiple concept levels,'' in \emph{Proceedings of the fourth international conference on Information and knowledge management}, 1995, pp. 19--24.

\bibitem{sharma2019brief}
D.~Sharma, R.~Shukla, A.~K. Giri, and S.~Kumar, ``A brief review on search engine optimization,'' in \emph{2019 9th international conference on cloud computing, data science \& engineering (confluence)}.\hskip 1em plus 0.5em minus 0.4em\relax IEEE, 2019, pp. 687--692.

\bibitem{jumper2021highly}
J.~Jumper, R.~Evans, A.~Pritzel, T.~Green, M.~Figurnov, O.~Ronneberger, K.~Tunyasuvunakool, R.~Bates, A.~{\v{Z}}{\'\i}dek, A.~Potapenko \emph{et~al.}, ``Highly accurate protein structure prediction with alphafold,'' \emph{nature}, vol. 596, no. 7873, pp. 583--589, 2021.

\bibitem{app14177782}
\BIBentryALTinterwordspacing
S.~Shahriar, B.~D. Lund, N.~R. Mannuru, M.~A. Arshad, K.~Hayawi, R.~V.~K. Bevara, A.~Mannuru, and L.~Batool, ``Putting gpt-4o to the sword: A comprehensive evaluation of language, vision, speech, and multimodal proficiency,'' \emph{Applied Sciences}, vol.~14, no.~17, 2024. [Online]. Available: \url{https://www.mdpi.com/2076-3417/14/17/7782}
\BIBentrySTDinterwordspacing

\bibitem{openai2023chatgptsearch}
OpenAI, ``Introducing chatgpt search,'' \url{https://openai.com/index/introducing-chatgpt-search/}, 2024, accessed: 2024-12-27.

\bibitem{sun2023chatgpt}
W.~Sun, L.~Yan, X.~Ma, S.~Wang, P.~Ren, Z.~Chen, D.~Yin, and Z.~Ren, ``Is chatgpt good at search? investigating large language models as re-ranking agents,'' \emph{arXiv preprint arXiv:2304.09542}, 2023.

\bibitem{team2023gemini}
G.~Team, R.~Anil, S.~Borgeaud, J.-B. Alayrac, J.~Yu, R.~Soricut, J.~Schalkwyk, A.~M. Dai, A.~Hauth, K.~Millican \emph{et~al.}, ``Gemini: a family of highly capable multimodal models,'' \emph{arXiv preprint arXiv:2312.11805}, 2023.

\bibitem{piasecki2018google}
J.~Piasecki, M.~Waligora, and V.~Dranseika, ``Google search as an additional source in systematic reviews,'' \emph{Science and engineering ethics}, vol.~24, pp. 809--810, 2018.

\bibitem{cilibrasi2007google}
R.~L. Cilibrasi and P.~M. Vitanyi, ``The google similarity distance,'' \emph{IEEE Transactions on knowledge and data engineering}, vol.~19, no.~3, pp. 370--383, 2007.

\bibitem{lin2005recall}
C.~Lin, ``Recall-oriented understudy for gisting evaluation (rouge),'' \emph{Retrieved August}, vol.~20, p. 2005, 2005.

\bibitem{papineni2002bleu}
K.~Papineni, S.~Roukos, T.~Ward, and W.-J. Zhu, ``Bleu: a method for automatic evaluation of machine translation,'' in \emph{Proceedings of the 40th annual meeting of the Association for Computational Linguistics}, 2002, pp. 311--318.

\bibitem{10.1371/journal.pone.0303231}
\BIBentryALTinterwordspacing
J.~K. Gill, M.~Chetty, S.~Lim, and J.~Hallinan, ``Large language model based framework for automated extraction of genetic interactions from unstructured data,'' \emph{PLOS ONE}, vol.~19, no.~5, pp. 1--22, 05 2024. [Online]. Available: \url{https://doi.org/10.1371/journal.pone.0303231}
\BIBentrySTDinterwordspacing

\bibitem{8665584}
M.~Ahmed, J.~Islam, M.~R. Samee, and R.~E. Mercer, ``Identifying protein-protein interaction using tree lstm and structured attention,'' in \emph{2019 IEEE 13th International Conference on Semantic Computing (ICSC)}, 2019, pp. 224--231.

\bibitem{10.1016/j.jbi.2019.103294}
\BIBentryALTinterwordspacing
Y.~Zhang, H.~Lin, Z.~Yang, J.~Wang, Y.~Sun, B.~Xu, and Z.~Zhao, ``Neural network-based approaches for biomedical relation classification: A review,'' \emph{J. of Biomedical Informatics}, vol.~99, no.~C, Nov. 2019. [Online]. Available: \url{https://doi.org/10.1016/j.jbi.2019.103294}
\BIBentrySTDinterwordspacing

\bibitem{lee2020biobert}
J.~Lee, W.~Yoon, S.~Kim, D.~Kim, S.~Kim, C.~H. So, and J.~Kang, ``Biobert: a pre-trained biomedical language representation model for biomedical text mining,'' \emph{Bioinformatics}, vol.~36, no.~4, pp. 1234--1240, 2020.

\bibitem{https://doi.org/10.1155/2014/298473}
\BIBentryALTinterwordspacing
D.~Zhou, D.~Zhong, and Y.~He, ``Biomedical relation extraction: From binary to complex,'' \emph{Computational and Mathematical Methods in Medicine}, vol. 2014, no.~1, p. 298473, 2014. [Online]. Available: \url{https://onlinelibrary.wiley.com/doi/abs/10.1155/2014/298473}
\BIBentrySTDinterwordspacing

\bibitem{sarmah2024hybridrag}
B.~Sarmah, D.~Mehta, B.~Hall, R.~Rao, S.~Patel, and S.~Pasquali, ``Hybridrag: Integrating knowledge graphs and vector retrieval augmented generation for efficient information extraction,'' in \emph{Proceedings of the 5th ACM International Conference on AI in Finance}, 2024, pp. 608--616.

\bibitem{Buehler_2024}
\BIBentryALTinterwordspacing
M.~J. Buehler, ``Accelerating scientific discovery with generative knowledge extraction, graph-based representation, and multimodal intelligent graph reasoning,'' \emph{Machine Learning: Science and Technology}, vol.~5, no.~3, p. 035083, sep 2024. [Online]. Available: \url{https://dx.doi.org/10.1088/2632-2153/ad7228}
\BIBentrySTDinterwordspacing

\bibitem{Zuluaga_Robledo_Arbelaez-Echeverri_Osorio-Zuluaga_Duque-Méndez_2022}
\BIBentryALTinterwordspacing
M.~Zuluaga, S.~Robledo, O.~Arbelaez-Echeverri, G.~A. Osorio-Zuluaga, and N.~Duque-Méndez, ``Tree of science - tos: A web-based tool for scientific literature recommendation. search less, research more!'' \emph{Issues in Science and Technology Librarianship}, no. 100, Aug. 2022. [Online]. Available: \url{https://journals.library.ualberta.ca/istl/index.php/istl/article/view/2696}
\BIBentrySTDinterwordspacing

\bibitem{10.1007/s10994-021-06068-6}
\BIBentryALTinterwordspacing
W.~Huang, X.~Zhao, and X.~Huang, ``Embedding and extraction of knowledge in tree ensemble classifiers,'' \emph{Mach. Learn.}, vol. 111, no.~5, p. 1925–1958, May 2022. [Online]. Available: \url{https://doi.org/10.1007/s10994-021-06068-6}
\BIBentrySTDinterwordspacing

\bibitem{yu2024multisourceknowledgepruningretrievalaugmented}
\BIBentryALTinterwordspacing
S.~Yu, M.~Cheng, J.~Yang, J.~Ouyang, Y.~Luo, C.~Lei, Q.~Liu, and E.~Chen, ``Multi-source knowledge pruning for retrieval-augmented generation: A benchmark and empirical study,'' 2024. [Online]. Available: \url{https://arxiv.org/abs/2409.13694}
\BIBentrySTDinterwordspacing

\bibitem{fan2024workflowllmenhancingworkfloworchestration}
\BIBentryALTinterwordspacing
S.~Fan, X.~Cong, Y.~Fu, Z.~Zhang, S.~Zhang, Y.~Liu, Y.~Wu, Y.~Lin, Z.~Liu, and M.~Sun, ``Workflowllm: Enhancing workflow orchestration capability of large language models,'' 2024. [Online]. Available: \url{https://arxiv.org/abs/2411.05451}
\BIBentrySTDinterwordspacing

\bibitem{10.1007/s11837-021-04902-9}
\BIBentryALTinterwordspacing
Z.~Hong, L.~Ward, K.~Chard, B.~Blaiszik, and I.~Foster, ``Challenges and advances in information extraction from scientific literature: a review,'' \emph{JOM}, vol.~73, no.~11, pp. 3383--3400, 2021. [Online]. Available: \url{https://doi.org/10.1007/s11837-021-04902-9}
\BIBentrySTDinterwordspacing

\bibitem{xie2024bytesciencebridgingunstructuredscientific}
\BIBentryALTinterwordspacing
T.~Xie, H.~Zhang, S.~Wang, Y.~Wan, I.~Razzak, C.~Kit, W.~Zhang, and B.~Hoex, ``Bytescience: Bridging unstructured scientific literature and structured data with auto fine-tuned large language model in token granularity,'' 2024. [Online]. Available: \url{https://arxiv.org/abs/2411.12000}
\BIBentrySTDinterwordspacing

\end{thebibliography}

\end{document}